\documentclass[journal,12pt,draftclsnofoot,onecolumn,twoside]{IEEEtran}
\normalsize

\usepackage{lscape}
\usepackage{color,soul}
\usepackage{multirow}

\usepackage{ifpdf}
 \ifpdf
     \usepackage[pdftex]{graphicx}
 \else
     \usepackage[dvips]{graphicx}
 \fi

\graphicspath{{figures/}}

\newtheorem{proposition}{Property}

\usepackage{cite}
\usepackage{subfigure}

\usepackage{epstopdf}					
\usepackage[belowskip=-12pt,aboveskip=8pt,footnotesize]{caption}

\usepackage[cmex10]{amsmath}
\usepackage{amssymb}

\usepackage{algorithm}

\usepackage{algorithm}
\usepackage{algpseudocode}

\begin{document}
%

\title{{On the Identification of SM and Alamouti Coded SC-FDMA Signals: A Statistical-Based Approach}}

\author{\normalsize Yahia~A.~Eldemerdash,~\IEEEmembership{\normalsize  Member,~IEEE} and
        Octavia~A.~Dobre,~\IEEEmembership{\normalsize Senior Member,~IEEE}
\thanks{Yahia A. Eldemerdash and  Octavia A. Dobre are with the Faculty of Engineering and Applied Science, Memorial
University,  St. John's, Canada. 
Email: \{yahia.eldemerdash, odobre\}@mun.ca.}
}

\maketitle

\vspace{-0.6cm}
\begin{abstract}

Signal identification represents the task of a receiver to identify the signal type and its parameters, with applications to both military and commercial communications. In this paper, we investigate the identification of spatial multiplexing (SM) and Alamouti (AL) space-time block code (STBC) with single carrier frequency division multiple access (SC-FDMA) signals, when the receiver is equipped with a single antenna. We develop a discriminating  feature  based on a fourth-order statistic of the received signal, as well as a constant false alarm rate decision criterion which relies on the statistical properties of the feature estimate. Furthermore, we present the theoretical performance  analysis of the proposed identification algorithm. The algorithm does not require channel or noise power estimation, modulation classification, and block synchronization. Simulation results show the validity of the proposed algorithm, as well as a very good agreement with the theoretical analysis.

\end{abstract}


\vspace{-0.45cm}
\section{INTRODUCTION}

Signal identification represents a core process in  military communications, such as in radio surveillance, electronic warfare, and  interference identification \cite{dobre2007survey,hameed2009likelihood,dobre2005blind}. For example, if the signal type is identified at the receiver, a more efficient jamming can be performed.  Recently, signal identification has attracted increased attention in several  commercial applications, such as software-defined and cognitive radios \cite{Dobre2015signal}.

Extensive studies have been devoted to signal identification problems for single-input single-output  scenarios (see the extensive survey \cite{dobre2007survey} and references therein). However, with the rapid adoption of the multiple-input multiple-output (MIMO) technology, additional signal identification problems have been introduced, such as estimation of the number of transmit antennas and identification of the  space-time block codes (STBCs).  Investigation of MIMO signal identification is at an early stage; for example, estimation of the number of transmit antennas was explored in \cite{hassan2011blind,ohlmer2008algorithm} and STBC identification was studied in \cite{choqueuse2010blind,choqueuse2008hierarchical,mohammadkarimi2014blind, eldemerdash2014blind, karami2014identification}.
For STBC identification, algorithms were developed for single carrier \cite{choqueuse2010blind,choqueuse2008hierarchical,mohammadkarimi2014blind} and orthogonal frequency division multiplexing (OFDM) systems \cite{eldemerdash2014blind,karami2014identification}. For the latter, receivers equipped with multiple antennas were considered, which is not always a practical case due to power, cost, and size limitations. Second-order statistics \cite{eldemerdash2014blind} and cyclic statistics \cite{karami2014identification} were employed as discriminating features; however, they are not applicable for the single receive antenna scenario. 
Furthermore, to the best of the authors' knowledge, there exists no study in the literature for the identification of single carrier frequency division multiple access (SC-FDMA) signals, which represent an alternative to OFDM, being used in uplink LTE and LTE-A \cite{ghosh2011essentials}.

This paper fills in the gap regarding the identification of STBC for SC-FDMA signals. Different from the existing works on OFDM signal identification, we propose
a fourth-order statistic-based algorithm to identify spatial multiplexing (SM) and Alamouti (AL) STBC for SC-FDMA signals when the receiver is equipped with a single antenna. The proposed fourth-order statistic exhibits significant peaks at certain delays for the AL SC-FDMA signal, whereas no peaks exist for the SM SC-FDMA signal.
A constant false alarm rate test is developed to detect the presence of such discriminating peaks. A theoretical analysis of the identification performance is performed and a closed form expression for the probability of correct identification is obtained. The proposed algorithm provides a good identification performance without requiring STBC/SC-FDMA block synchronization or knowledge of the modulation, channel, and noise power.

The rest of this paper is organized as follows. The system model is introduced in Section \ref{sec:Signal model_ch4}, and the properties of AL and SM SC-FDMA signals, leading to the discriminating feature, are presented in Section \ref{sec:Feature_ch4}.  
The proposed identification algorithm is described in Section \ref{sec:Alg_ch4}, and the corresponding  theoretical performance analysis is given in Section \ref{sec:Theortical_performnace_ch4}. Finally, simulation  
results are reported in Section \ref{sec:Simulation_ch4}, and conclusions are drawn in Section \ref{sec:conclusion_ch4}.

\section{System model \label{sec:Signal model_ch4}}

\begin{figure}
\begin{centering}
\includegraphics[width=0.85\textwidth]{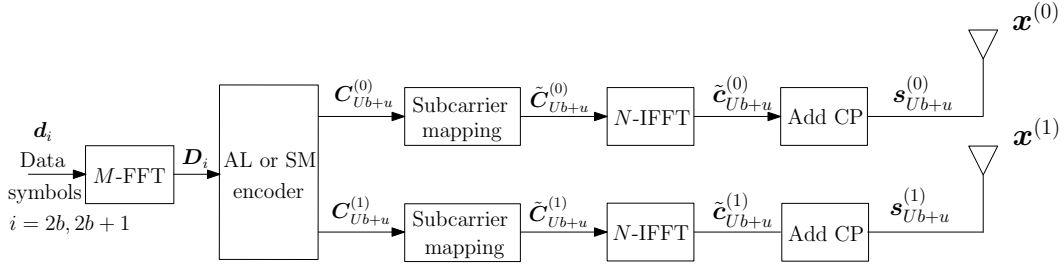}
\par\end{centering}
\caption{Block diagram of an AL/ SM SC-FDMA transmitter \cite{myung2008single}.\label{fig:Block_ch4}}
\end{figure}

The block diagram of a transmitter using either AL or SM SC-FDMA signals and equipped with two  antennas is shown in Fig. \ref{fig:Block_ch4}. The  unit-variance modulated data symbols, which are randomly and independently generated from an $\Omega$-point constellation, $\Omega \geq 4$, are considered as a stream of $M$-length blocks, i.e., the $i$th data block is $\boldsymbol{d}_i= [d_i(0),...,d_i(M-1)]$. Each block is input to an $M$-point fast Fourier transform (FFT) leading to a stream of $M$-length blocks representing the frequency-domain symbols corresponding to the modulated data blocks, $\boldsymbol{D}_i=[D_i(0),...,D_i(M-1)]$. Each two consecutive frequency-domain blocks, i.e., with $i=2b,2b+1$, are encoded using a space-time coding matrix which is respectively defined for AL and SM STBCs  as

%
%
\begin{equation}
\mathbb{G}^{(\textrm{AL})}=\left[\begin{array}{cc}
\boldsymbol{D}_{2b+0} & -\boldsymbol{D}_{2b+1}^{*}\\
\boldsymbol{D}_{2b+1} & \boldsymbol{D}_{2b+0}^{*}
\end{array}\right]=\left[\begin{array}{cc}
\boldsymbol{C}_{2b+0}^{(0)} & \boldsymbol{C}_{2b+1}^{(0)}\\
\boldsymbol{C}_{2b+0}^{(1)} & \boldsymbol{C}_{2b+1}^{(1)}
\end{array}\right],\label{eq:AL_SC_FDMA}
\end{equation}
and 
\begin{equation}
\mathbb{G}^{(\textrm{SM})}=\left[\begin{array}{c}
\boldsymbol{D}_{2b+0}\\
\boldsymbol{D}_{2b+1}
\end{array}\right]=\left[\begin{array}{c}
\boldsymbol{C}_{b+0}^{(0)}\\
\boldsymbol{C}_{b+0}^{(1)}
\end{array}\right],\label{eq:SM_SC_FDMA}
\end{equation}
where $\boldsymbol{C}^{(f)}_{Ub+u}$ represents the $(Ub+u)$th block of length $M$ transmitted  from the $f$th antenna, $f=0,1$ and $b$ is an integer denoting the STBC block index, $U$ is the length of the STBC ($U=2$ for AL and $U=1$ for SM), $u$ is the slot index within the $b$th STBC block, $u=0,...,U-1$, and * denotes the complex conjugate.   

For each transmit branch, the interleaved subcarrier mapping\footnote{Interleaved subcarrier mapping is considered to simplify the presentation of the discriminating feature. However, the proposed algorithm is applicable to localized mapping, as well; simulation results are presented in Section \ref{sec:Simulation_ch4}.} is used to assign the $M$ symbols within the block $\boldsymbol{C}^{(f)}_{Ub+u}$ to $N$ subcarriers, $N>M$. Note that $N=\lambda M$, with $\lambda$ as an integer representing the expansion factor, and the unoccupied subcarriers are set to zero \cite{ghosh2011essentials}. The mapped block  $\tilde{\boldsymbol{C}}^{(f)}_{Ub+u}$ is converted into a time-domain sequence  $\tilde{\boldsymbol{c}}^{(f)}_{Ub+u}=[\tilde{c}^{(f)}_{Ub+u}(0),..., \tilde{c}^{(f)}_{Ub+u}(N-1)]$ using an $N$-point inverse FFT (IFFT). Then, the cyclic prefix of length $\nu$ is added to the block $\tilde{\boldsymbol c}^{(f)}_{Ub+u}$ by appending the last $\nu$ samples as a prefix, leading to  $\boldsymbol{s}^{(f)}_{Ub+u} \hspace{-0.1cm} = \hspace{-0.1cm} [\tilde{c}^{(f)}_{Ub+u}(N\hspace{-0.1cm}-\hspace{-0.1cm}\nu),..., \tilde{c}^{(f)}_{Ub+u}(N-1), \hspace{-0.05cm}\tilde{c}^{(f)}_{Ub+u}(0),...,\hspace{-0.05cm} \tilde{c}^{(f)}_{Ub+u}(N-1)]$, with components expressed~as

\vspace{-0.35cm}
\begin{equation}
\begin{array}{l}
\hspace{-0.57cm}s_{Ub+u}^{(f)}(n)\hspace{-0.1cm}=\hspace{-0.1cm}\frac{1}{\sqrt{N}} \hspace{-0.1cm} \displaystyle \sum_{p=0}^{N-1} \hspace{-0.1cm} \tilde{C}_{Ub+u}^{(f)}(p)e^{\frac{j2\pi p(n-\nu)}{N}}, 
 n\hspace{-0.1cm} =\hspace{-0.1cm} 0,...,N\hspace{-0.1cm}+\hspace{-0.1cm}\nu \hspace{-0.1cm} -\hspace{-0.1cm} 1. 
 \label{eq:IFFT_eq_ch4}
\end{array}
\end{equation}

Accordingly, the transmitted sequence from the $f$th antenna can be expressed as $\boldsymbol{x}^{(f)}=[\boldsymbol{s}^{(f)}_{0}, \boldsymbol{s}^{(f)}_{1}, ...]$. We consider a receiver equipped with a single antenna, where the $k$th  sample can be expressed~as

\vspace{-0.25cm}
\begin{equation}
y(k)=\boldsymbol{h} \boldsymbol{x}^T_k+w(k), \label{eq:RX_signal_ch4}
\end{equation}
where the superscript $T$ denotes  transpose, $\boldsymbol{h}=[h_0(0),...,h_0(L_h-1),h_1(0),..., h_1(L_h-1)]$,  $\boldsymbol{x}_k=[x^{(0)}(k),...,x^{(0)}(k-L_h-1), x^{(1)}(k),...,x^{(1)}(k-L_h-1)]$, 
with $L_h$ as the number of propagation paths, $h_{f}(l)$ as the channel coefficient corresponding to the $l$th path between the transmit antenna $f$ and the receive antenna, $l=0,...,L_h-1$, and $x^{(f)}(k-l)$ as the $(k-l)$th element of the sequence transmitted via the $f$th antenna, and $w(k)$ represents the zero-mean complex additive white Gaussian noise (AWGN) with  variance~$\sigma^2_w$.

\vspace{-0.25cm}
\section{Properties of AL and SM SC-FDMA signals and discriminating feature \label{sec:Feature_ch4}}

In this section, we first introduce  properties exhibited by the AL and SM SC-FDMA signals, which we
 then employ to obtain the feature for their identification.



\begin{proposition}
One can show that the time-domain samples $\tilde{{c}}^{(0)}_{2b+0}(v)$ and $\tilde{{c}}^{(1)}_{2b+1}(v)$, $v=0,...,N-1$, which correspond to ${\boldsymbol{C}}^{(0)}_{2b+0}={\boldsymbol{D}}_{2b+0}\textendash \textendash$see the AL encoding matrix in (1) for $u=0$ and $f=0 \textendash \textendash$ and ${\boldsymbol{C}}^{(1)}_{2b+1}={\boldsymbol{D}}^*_{2b+0} \textendash \textendash$see the  AL encoding matrix in (1) for $u=1$ and $f=1 \textendash \textendash$are expressed as a function of their associated modulated data symbols, $\boldsymbol{d}_{2b+0}$ and $\boldsymbol{d}^*_{2b+0}$, respectively as

\vspace{-0.25cm}
\begin{equation}
\begin{array}{l}
\tilde{{c}}^{(0)}_{2b+0}(v=m+aM)=\frac{1}{\sqrt{\lambda}} d_{2b+0}(m),
\qquad \qquad m=0,...,M-1, a=0,...,\lambda-1,
\end{array} \label{eq:d_1_ch4}
\end{equation}


\vspace{-0.15cm}

\begin{equation}
\begin{array}{l}
\tilde{{c}}^{(1)}_{2b+1}(v=m+aM)=\frac{1}{\sqrt{\lambda}} d_{2b+0}^*(mod(-m,M)),
\qquad \qquad m=0,...,M-1, a=0,...,\lambda-1,
\end{array}\label{eq:d_2_ch4}
\end{equation}
where $mod$ is the modulo operation. Similar expressions can be written for $\tilde{{c}}^{(1)}_{2b+0}(v)$ and $\tilde{{c}}^{(0)}_{2b+1}(v)$, $v=0,...,N-1$, which correspond to ${\boldsymbol{C}}^{(1)}_{2b+0}={\boldsymbol{D}}_{2b+1}$ and ${\boldsymbol{C}}^{(0)}_{2b+1}={-\boldsymbol{D}}^*_{2b+1}$, as a function of their associated modulated data symbols, $\boldsymbol{d}_{2b+1}$ and $\boldsymbol{d}^*_{2b+1}$, respectively.  
\end{proposition}


For SM, similar to (\ref{eq:d_1_ch4}), the time domain samples transmitted from the two antennas can be expressed as $\tilde{{c}}^{(0)}_{b+0}(v=m+aM)=\frac{1}{\sqrt{\lambda}} d_{2b+0}(m)$ and $\tilde{{c}}^{(1)}_{b+0}(v=m+aM)=\frac{1}{\sqrt{\lambda}} d_{2b+1}(m)$, $m=0,...,M-1$, $a=0,...,\lambda-1$.

In other words, the transmitted symbols from each antenna belong to the $\Omega$-point constellation, scaled by a factor of $\frac{1}{\sqrt{\lambda}}$; note that this also holds after adding the cyclic prefix.

\begin{proposition}
Based on the structure of the AL coding matrix and by following \cite{eldemerdash2014blind}, one can  show for the AL SC-FDMA signal that $s_{2b+0}^{(0)}(n)=s_{2b+1}^{(1)^{*}}(mod(-(n- \nu),N)+ \nu)$,   $\forall n\hspace{-0.1cm}=\hspace{-0.1cm}0,...,N+\nu-1.$


\end{proposition}

\begin{proposition}
Let $\boldsymbol{x}^{(f,\tau)}$ denote the sequence with components  $x^{(f,\tau)}(k)=x^{(f)}(k+\tau)$, $\tau=0,...,N+\nu-1$. This is split into consecutive $(N+\nu)$-length blocks, i.e., $\boldsymbol{x}^{(f,\tau)}= [\boldsymbol{{s}}^{(f,\tau)}_{0}, \boldsymbol{{s}}^{(f,\tau)}_{1}, ...,$ $\boldsymbol{{s}}^{(f,\tau)}_{i-1}, \boldsymbol{{s}}^{(f,\tau)}_{i}$, $\boldsymbol{{s}}^{(f,\tau)}_{i+1},...]$.  
By using \textit{Property 1} and \textit{Property 2}, and following the analysis in \cite{eldemerdash2014blind}, for the AL SC-FDMA signal, it can be shown that the components of $\boldsymbol{s}_i^{(0,\tau)}$ and $\boldsymbol{s}_{i+1}^{(1,\tau)}$ satisfy\footnote{Note that $\lambda=2$ is considered in the analysis,   as being a typical value~\cite{myung2008single}.}

\vspace{-0.2cm}
\begin{equation}
s_{i}^{(0,\tau)}(n)=s_{i+1}^{(1,\tau)^{*}}(mod(-(n- \nu),N)+ \nu),
\label{eq:G_0_ch4}
\end{equation}
if and only if: $ \tau=0$,  $i=2b$, and $n=0,...,N+\nu-1$; $ \tau=N/4$,  $i=2b$, and $n=0,...,\nu$, $n=N/4+\nu+1,...,3N/4+\nu-1$;  $  \tau=N/2$, $i=2b$, and $n=0,...,\nu$;  
 $ \tau=N/2+\nu$, $i=2b-1$, and $n=\frac{N}{2},...,\frac{N}{2}+2\nu$;  
  $\tau=3N/4$,  $i=2b$, and $n=0,...,\nu$; $\tau=3N/4+\nu$,  $i=2b-1$, and $n=N/4,...,3N/4+2\nu$.

%
%
%
%
%

\end{proposition}


%
%

%
%
%

 We first introduce the proposed fourth-order statistic of the transmitted sequence, and then the one of the received sequence.  We define the fourth-order statistic of the transmitted sequence as

\vspace{-0.45cm}
\begin{equation}
\mathcal{A}_s(\tau)=\textrm{E}\left\{ \left[\boldsymbol{s}_{i}^{(0,\tau)} \circ \boldsymbol{s}_{i}^{(0,\tau)} \right]\left[\bar{\boldsymbol{s}}_{i+1}^{(1,\tau)}\circ \bar{\boldsymbol{s}}_{i+1}^{(1,\tau)} \right]^{T}\right\}, \label{eq:A_tr_ch4}
\end{equation}
where $\boldsymbol{\bar{s}}_{i+1}^{(1,\tau)}=[\bar{s}_{i+1}^{(1,\tau)}(0), . . ., \bar{s}_{i+1}^{(1,\tau)}(N+\nu-1)]$, with $\bar{s}_{i+1}^{(1,\tau)}(n)= s_{i+1}^{(1,\tau)}(mod(-(n- \nu),N)+ \nu)$,  $n=0,...,N+\nu-1$, and $\textrm{E}$ and $\circ$  denote the statistical expectation and the  Hadmard product, respectively.   

Based on  (\ref{eq:IFFT_eq_ch4}), \textit{Property 1},  and \textit{Property 3}, one can show that $\mathcal{A}_s^{\textrm{AL}}(\tau)=\frac{\kappa_{d,4,2}}{4}(N+\nu)$ for $\tau=0$;
$\frac{\kappa_{d,4,2}}{4}(\frac{N}{2}+\nu)$ for $\tau=N/4$; 
$\frac{\kappa_{d,4,2}}{4}(\nu+1)$ for $\tau=N/2$; 
$\frac{\kappa_{d,4,2}}{4}(2\nu+1)$ for $\tau=N/2+\nu$;  $\frac{\kappa_{d,4,2}}{4}(\nu+1)$ for $\tau=3N/4$; $\frac{\kappa_{d,4,2}}{4}(\frac{N}{2}+2\nu+1)$ for $\tau=3N/4+\nu$; and $0$, otherwise.
%
%
Here  $\kappa_{d,4,2}$ represents the  (4,2) cumulant \cite{dobre2007survey} of the modulated data symbols. 
On the other hand, it can be easily shown that $\mathcal{A}_s^{\textrm{SM}}(\tau)=0$, $\forall \tau=0,...,N+\nu-1.$

At the receive-side,  the following assumptions and definitions are first introduced:

\hspace{-0.55cm} 1) Without loss of generality, we assume that the first intercepted sample corresponds to the start of an SC-FDMA block, and the total number of  received samples is a multiple integer of the SC-FDMA block length, i.e., $K=N_B(N+\nu)$, where $N_B$ is the number of received SC-FDMA blocks. This assumption will be relaxed later in the paper.

\hspace{-0.55cm} 2) Define the vector $\boldsymbol{y}^{(\tau)}$ with $y^{(\tau)}(k)=y(k+\tau)$, $\tau=0,...,N+\nu-1$. This vector is then divided into consecutive blocks each of length $(N+\nu)$,\footnote{ Note that  the block length, $(N+\nu)$, is assumed known at the receive-side; otherwise, this parameter needs to be estimated. For example, the algorithm in \cite{zhang2013second}, for single transmit antenna scenario, can be adapted to the scenario under consideration to estimate  the block length.} i.e.,
$\boldsymbol{y}^{(\tau)}=[ \boldsymbol{r}_{0}^{(\tau)},...,\boldsymbol{r}_{N_{B}-1}^{(\tau)}] $,
where $\boldsymbol{r}_{q}^{(\tau)}=[r_q^{(\tau)}(0), . . . ,r_q^{(\tau)}(N+\nu-1)]$, with $r_{q}^{(\tau)}(n)=y^{(\tau)}(n+q(N+\nu)),\; n=0,...,N+\nu-1,  \;  q=0,...,N_B-1.$ 

%
%

\hspace{-0.55cm} 3) Define the fourth-order statistic $\mathcal{A}_r(\tau)\hspace{-0.1cm}=\hspace{-0.1cm} \textrm{E}\Bigl\{ \left[\boldsymbol{r}_{q}^{(\tau)} \circ \boldsymbol{r}_{q}^{(\tau)} \right]$ $\left[\bar{\boldsymbol{r}}_{q+1}^{(\tau)}\circ \bar{\boldsymbol{r}}_{q+1}^{(\tau)} \right]^{T}\Bigr\}$, where $\boldsymbol{\bar{r}}_{q+1}^{(\tau)}\hspace{-0.1cm} = \hspace{-0.1cm} $ $[\bar{r}_{q+1}^{(\tau)}(0),$ $ . . ., $ $\bar{r}_{q+1}^{(\tau)}(N+\nu-1)]$, with $\bar{r}_{q+1}^{(\tau)}(n) \hspace{-0.1cm}= \hspace{-0.1cm}r_{q+1}^{(\tau)}(mod(-(n- \nu),N)+ \nu)$, $n=0,...,N+\nu-1$.


By using (\ref{eq:RX_signal_ch4}) and (\ref{eq:G_0_ch4}),  one can obtain $\mathcal{A}_r (\tau)$ for the AL SC-FDMA signal as $\mathcal{A}_r^{\textrm{AL}}(\tau)=(N+\nu) \Psi (\tau)$ for $ \tau=0,...,L_h-1$;
$(\frac{N}{2}+\nu)\Psi (\tau-\frac{N}{4})$ for  $\tau=\frac{N}{4}, ..., \frac{N}{4}+L_h-1$;
$(\nu+1)\Psi (\tau-\frac{N}{2})$ for $\tau=\frac{N}{2}, ...,\frac{N}{2}+L_h-1$; $(2\nu+1)\Psi (\tau-\frac{N}{2}-\nu)$ for $\tau=\frac{N}{2}+\nu, ..., \frac{N}{2}+\nu+L_h-1$; 
$(\nu+1)\Psi (\tau-\frac{3N}{4})$ for $\tau=\frac{3N}{4}, ...,\frac{3N}{4}+L_h-1$; 
$(\frac{N}{2}+2\nu+1)\Psi (\tau-\frac{3N}{4}-\nu)$ for $ \tau=\frac{3N}{4}+\nu, ..., \frac{3N}{4}+\nu+L_h-1$; $0$, otherwise.
Here $\Psi (\tau) =\frac{\kappa_{d,4,2}}{4} \sum_{l,l'=0}^{L_h-1} (h_0^2(l) h_1^2(l')) \delta(\tau-l-l')$.

On the other hand, by using the properties that $\textrm{E} \{ d_i^2(m) \}=\textrm{E} \{ d_i^4(m) \}=0$ for the symbols drawn from $\Omega$-point constellation, $\Omega \geq 4$ \cite{dobre2007survey}, and that the signal, noise, and channel coefficients are independent,
one can  easily show that $\mathcal{A}_r^{\textrm{SM}}(\tau)=0$, $\forall \tau=0,...,N+\nu-1.$


\begin{figure}
\begin{centering}
\includegraphics[width=0.7\textwidth]{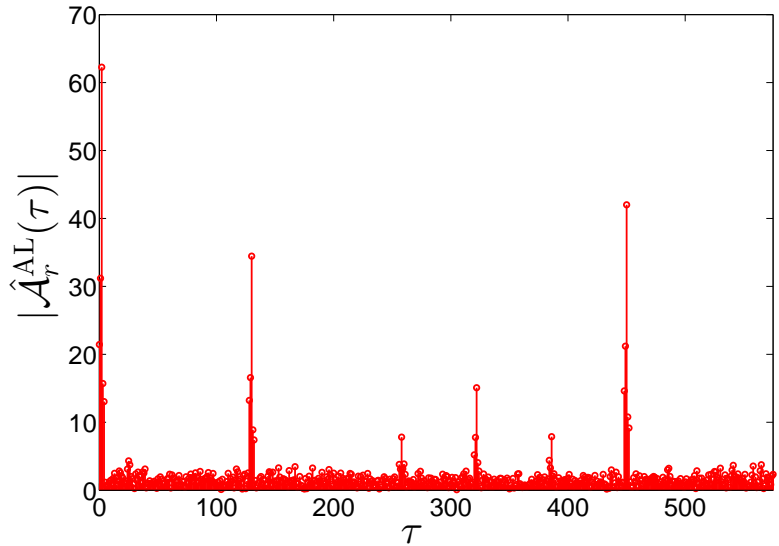}
\par\end{centering}
\caption{ $| \hat{\mathcal{A}}_r^{\textrm{AL}} (\tau)|$ with QPSK modulation,  $N=512$, $\lambda=2$, $\nu=N/8$, and $N_B=4000$ at SNR= 25 dB over multipath Rayleigh fading channel, $L_h=3$, for the AL SC-FDMA signal.\label{fig:peaks_ch4}}
\end{figure}

Fig. \ref{fig:peaks_ch4} shows the magnitude of the estimated fourth-order statistic, $|\hat{\mathcal{A}}_r^{\textrm{AL}} (\tau)|$, with $N=512$, $M=N/2$,  $\nu=N/8$, $N_B=4000$, and quadrature phase-shift-keying (QPSK) modulation over multipath Rayleigh fading channel with $L_h=3$ at SNR= $25$ dB. Apparently, the results match with the theoretical findings discussed above.
It is worth noting that if the first received sample does not correspond to the start of an SC-FDMA block, the discriminating peaks shown in Fig. \ref{fig:peaks_ch4} are shifted with the delay between the first received sample and the start of the next SC-FDMA block. Indeed, this shift does not affect the identification performance, as the statistically significant peaks still exist and they maintain their values.

\vspace{-0.3cm}
\section{Proposed identification algorithm \label{sec:Alg_ch4}}

Based on the aforementioned discussion, the presence of significant peaks in $|\hat{\mathcal{A}}^{\textrm{AL}}_r(\tau)|$ provides a discriminating feature that can be employed to identify AL and SM SC-FDMA signals. Here we propose a statistical test to detect the peak presence in $|\hat{\mathcal{A}_r}(\tau)|$, and develop the identification algorithm accordingly. The fourth-order statistic  can be estimated as

\vspace{-0.45cm}
\begin{equation}
\hat{\mathcal{A}_r}(\tau)={N_{B}^{-1}}\sum_{q=0}^{N_{B}-1}\left[\boldsymbol{r}_{q}^{(\tau)} \circ \boldsymbol{r}_{q}^{(\tau)} \right]\left[\bar{\boldsymbol{r}}_{q+1}^{(\tau)}\circ \bar{\boldsymbol{r}}_{q+1}^{(\tau)} \right]^{T}.
\label{eq:Rm_est_ch4}\end{equation}

By following \cite{kay1993fundamentals}, this can be expressed as $\hat{\mathcal{A}_r}(\tau)=\mathcal{A}_r(\tau)+\epsilon(\tau)$,
%
%
%
where  the estimation error $\epsilon(\tau)$ has an asymptotic zero-mean complex Gaussian distribution with variance $\sigma^2_{\epsilon}$. For SM SC-FDMA (hypothesis $\mathcal{H}_0$), $\mathcal{A}_r(\tau)=0$ $\forall \tau=0,...,N+\nu-1$, while for AL SC-FDMA (hypothesis $\mathcal{H}_1$)  $\mathcal{A}_r(\tau) \neq 0$. We define the statistic $\mathcal{G}(\tau)$ as



\vspace{-0.45cm}
\begin{equation}
\mathcal{G}(\tau)=\frac{2|\hat{\mathcal{A}_r}(\tau)|^{2}}{{(N+\nu)^{-1}}{\displaystyle \sum^{N+\nu-1}_{\tau'=0} |\hat{\mathcal{A}}'_r(\tau')|^{2}}},  \tau=0,...,N+\nu-1,\label{eq:Test_function_ch4}
\end{equation}
\vspace{-0.2cm}
with
\vspace{-0.2cm}
\begin{equation}
{\hat{\mathcal{A}}'_r}(\tau')={N_{B}^{-1}}\sum_{q=0}^{N_{B}-1}\left[\boldsymbol{r}_{q}^{(\tau')} \circ \boldsymbol{r}_{q}^{(\tau')} \right]\left[\bar{\boldsymbol{r}}_{q+4}^{(\tau')}\circ \bar{\boldsymbol{r}}_{q+4}^{(\tau')} \right]^{T}.
\label{eq:sigma_est_ch4}\end{equation}

Note that the denominator in (\ref{eq:Test_function_ch4}) is an estimate of $\sigma_{\epsilon}^2$ regardless of the transmitted signal (there are no statistically significant peaks). Also, for the AL SC-FDMA signal, the peak positions in $\mathcal{G} (\tau)$ are the same as in $\mathcal{A}_r^{\textrm{AL}} (\tau)$. Moreover, for the SM SC-FDMA signal,  $\hat{\mathcal{A}_r}(\tau)=\epsilon(\tau)$ has an asymptotic zero-mean complex Gaussian distribution with variance $\sigma_{\epsilon}^2$. Accordingly, and considering the denominator of (\ref{eq:Test_function_ch4}) as a constant, $\sigma^2_\epsilon$, $\mathcal{G}(\tau)$ has an asymptotic central chi-square distribution with the degree of freedom equal to two \cite{kay1993fundamentals}. 
We further define the test statistic  $\Gamma=\max_{\tau=0,...,N+\nu-1} \mathcal{G}(\tau)$.


By using the fact that the cumulative distribution function (CDF) of the maximum value $\Gamma$ is the product of the CDFs of $\mathcal{G}(\tau)$, $\tau=0,...,N+\nu-1$ \cite{kay1993fundamentals},  we can set a threshold, $\gamma$ corresponding to a certain probability of false alarm, $P_{f}=P(\Gamma > \gamma|\mathcal{H}_0 )$, i.e., $1-P_{f}=(1-e^{\frac{-\gamma}{2}})^{(N+\nu)}$.
Then, the threshold can be calculated as $\gamma= -2 \ln (1-(1-P_{f})^{\frac{1}{N+\nu}})$.




\vspace{-0.25cm}
\section{Theoretical Performance analysis \label{sec:Theortical_performnace_ch4}}
\vspace{-0.15cm}
The identification process is done by comparing the test statistic, $\Gamma$, with a threshold, $\gamma$. If $\Gamma> \gamma$, then the AL SC-FDMA signal is declared present; otherwise, SM SC-FDMA is chosen. As the threshold $\gamma$ is determined according to a constant $P_{f}$, the probability of correctly identifying the SM SC-FDMA signal is determined as $P(\zeta = \textrm{SM} | \textrm{SM}) = 1- P_{f}$,
%
where $\zeta$ is the estimated signal type.

On the other hand, the probability of correctly identifying the AL SC-FDMA signal is $P(\zeta = \textrm{AL} | \textrm{AL})= P(\Gamma > \gamma | \mathcal{H}_1)=1-P(\Gamma \leq \gamma | \mathcal{H}_1)$. As  $\Gamma$ is the maximum of $\mathcal{G}(\tau)$, $P(\Gamma \leq \gamma | \mathcal{H}_1)$ is the probability that $\mathcal{G}(\tau) \leq \gamma$, $\forall \tau=0,...,N+\nu-1$.  Furthermore, under hypothesis $\mathcal{H}_1$, $\mathcal{G} (\tau)$ has significant peaks around $\tau=0, N/4, N/2, N/2+\nu, 3N/4, 3N/4+\nu$  and has nulls at other values of $\tau$.  Let $\Lambda$ be the set of $\tau$ values for which $\mathcal{G} (\tau)$ is non-zero.  According to the expression of $\mathcal{A}_r^{\textrm {AL}}(\tau)$, the cardinality of the set $\Lambda$ is $6L_h$; as such, $N+\nu-6L_h$ points of $\mathcal{G}(\tau),\: \tau \notin \Lambda$, have an asymptotic central chi-square distribution with two degrees of freedom, and one can write
$P(\mathcal{G}(\tau) \leq \gamma, \tau \notin \Lambda | \mathcal{H}_1)=(1-e^{-\gamma/2})^{N+\nu-6L_h}$.
%
For  $\tau \in \Lambda$, the corresponding values of $\mathcal{G} (\tau)$ have a non-central chi-square distribution with the non-centrality parameter, $\mathcal{P}_\tau$,  which can be written based on (\ref{eq:Test_function_ch4}) as

\begin{figure*}
\subfigure[]{\includegraphics[width=.35\textwidth]{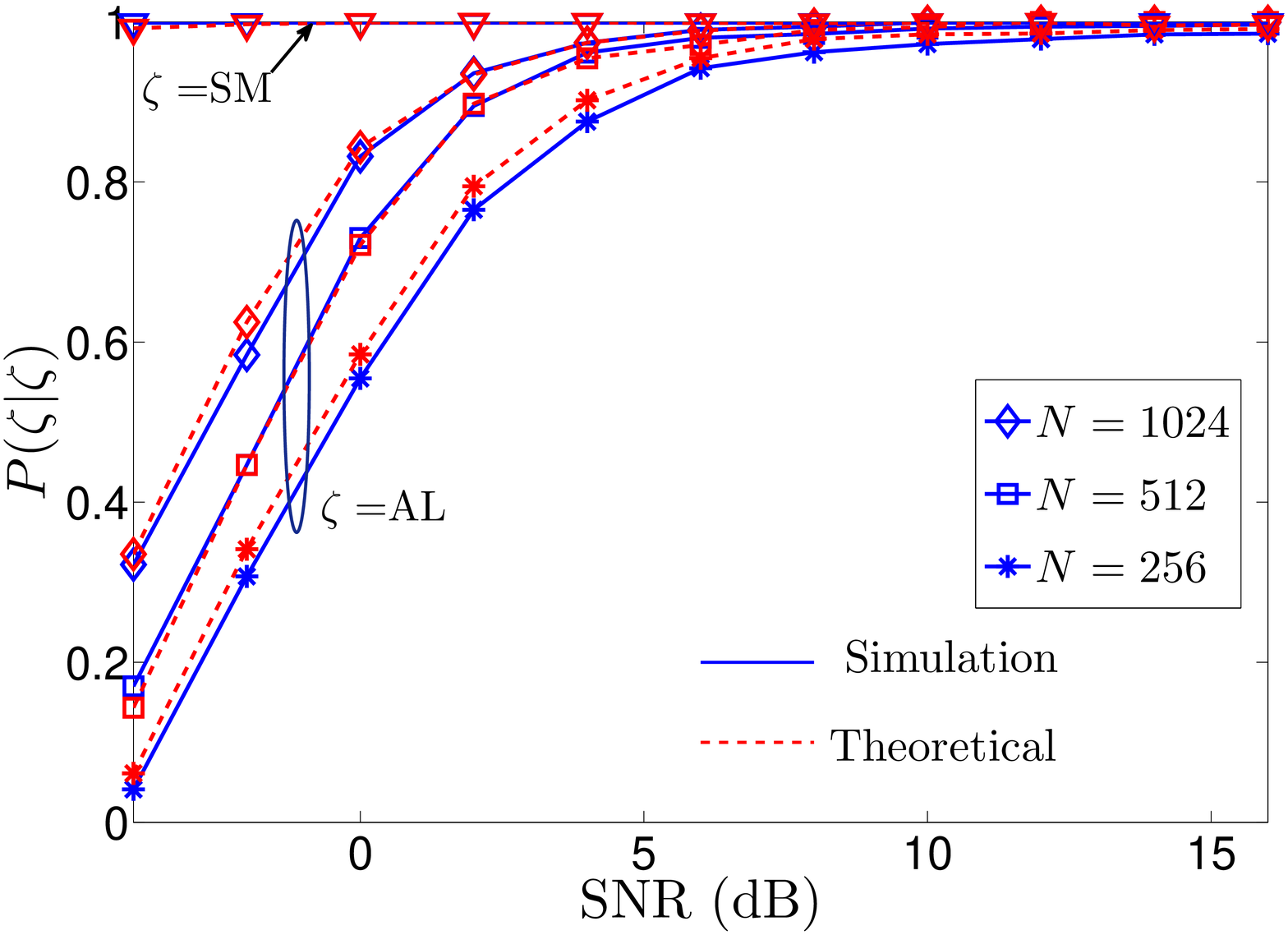}}\hspace{-0.65cm}
\subfigure[]{\includegraphics[width=.35\textwidth]{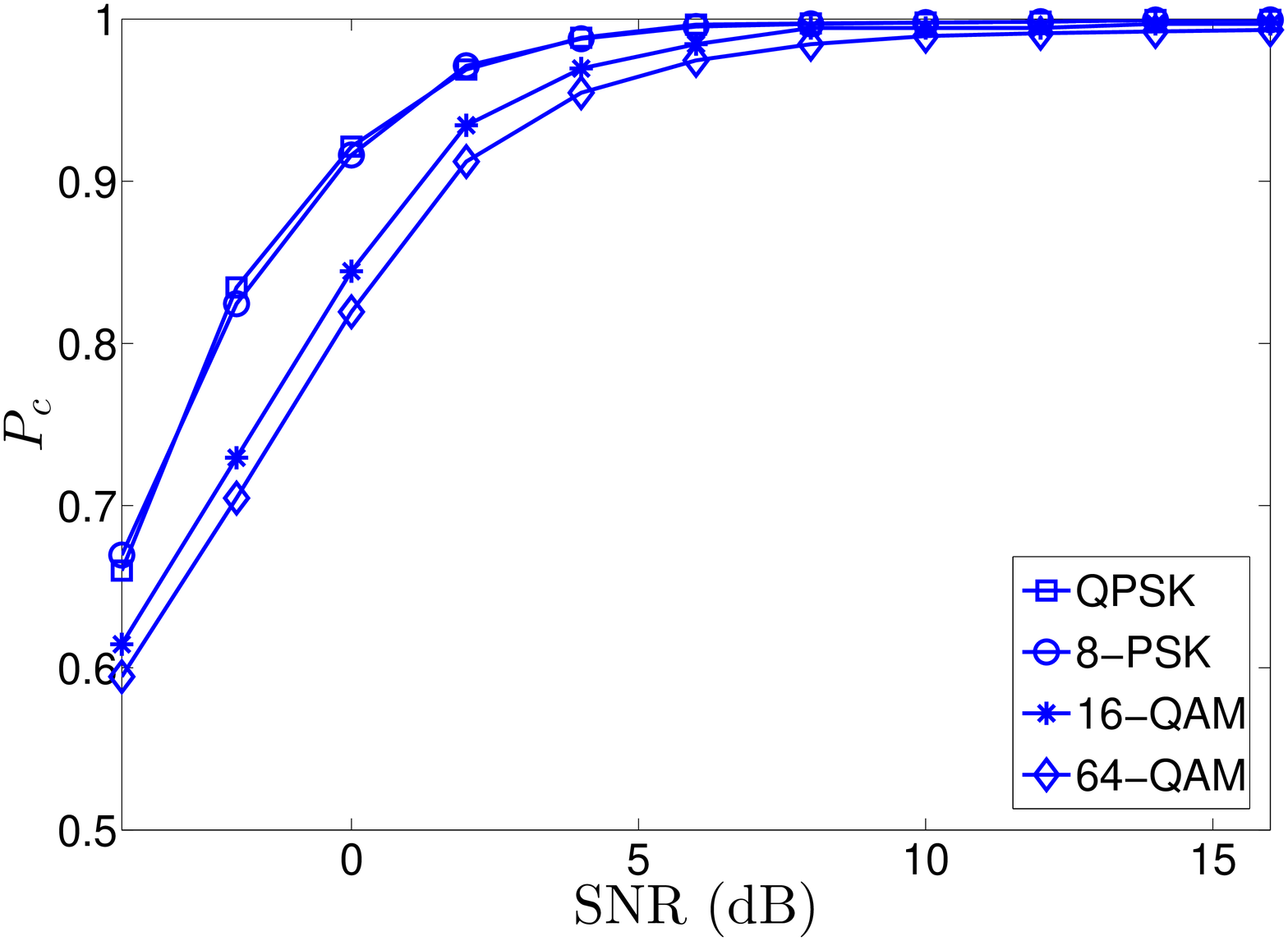}} \hspace{-0.65cm}
\subfigure[]{\includegraphics[width=.35\textwidth]{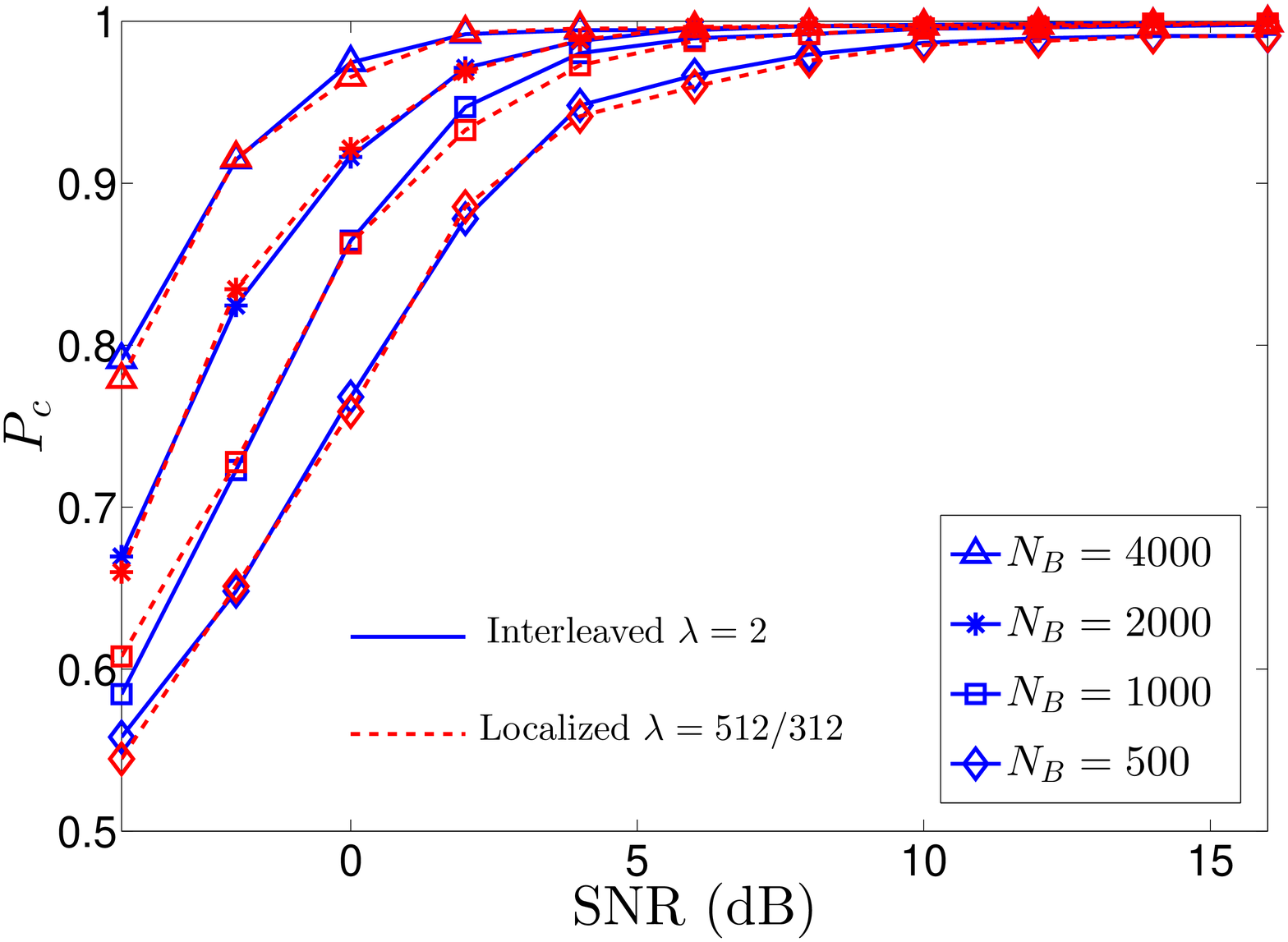}} \hspace{-0.65cm}
\caption{Performnace evaluation of the proposed algorithm: (a)  effect of the number of subcarriers, (b)  effect of the modulation format,   (c)  effect of subcarrier mapping. \label{fig:NB_ch4}}
\end{figure*}

\vspace{-0.55cm}
\begin{equation}
\mathcal{P}_\tau = \frac{2|\mathcal{A}_r(\tau)|^2}{\sigma^2_{\epsilon}}.
\end{equation}

For the PSK modulated data symbols with unit variance constellation,\footnote{Note that in this case both second-order/ one-conjugate and fourth-order/ two-conjugate moments are equal to one \cite{dobre2007survey}, and a simple analytical expression can be obtained.} by using that $\boldsymbol{r}_{q}^{(\tau)}$ and $\boldsymbol{r}_{q+4}^{(\tau)}$ are independent,  representing them as a function of $y^{(\tau)}(k)$, and employing that $y^{(\tau)}(k)=y(k+\tau)$, along with  (\ref{eq:RX_signal_ch4}), one can show that $\sigma^2_{\epsilon}$ is given by 


\vspace{-0.35cm}
\begin{equation}
\begin{array}{ll}
\hspace{-0.25cm}\sigma_{\epsilon}^2&\hspace{-0.3cm}= \hspace{-0.1cm}\frac{N+\nu}{N_B}\Big[ \frac{1}{\lambda^4}\parallel \hspace{-0.1cm}(\boldsymbol{h}\otimes \boldsymbol{h}) \otimes (\boldsymbol{h} \otimes \boldsymbol{h} )\hspace{-0.1cm} \parallel_F^2 + 8 \frac{\sigma^2_w}{\lambda^3} \parallel (\boldsymbol{h} \otimes \boldsymbol{h}) \otimes \boldsymbol{h} \parallel_F^2\\
& + 20 \frac{\sigma^4_w}{\lambda^2} \parallel \hspace{-0.1cm} (\boldsymbol{h} \otimes \boldsymbol{h})\hspace{-0.1cm} \parallel_F^2 
+16 \frac{\sigma_w^6}{\lambda} \parallel \boldsymbol{h} \parallel_F^2 + 4\sigma_w^8 \Big],
\end{array}\label{eq:var_th_ch4}
\end{equation}
where $\otimes$ and $\parallel \cdot \parallel_F$ denote the Kronecker product and the Frobenius norm, respectively.

Based on the CDF of the non-central chi-square distribution with two degrees of freedom, the probability that $\mathcal{G} (\tau) \leq \gamma$, $\tau \in \Lambda$, for  certain channel coefficients is 
$P(\mathcal{G}(\tau) \leq \gamma, \tau \in \Lambda | \mathcal{H}_1, \boldsymbol{h})= \prod_{\tau \in \Lambda} (1-Q_1 (\sqrt{\mathcal{P}_\tau}, \sqrt{\gamma}))$,
%
where $Q_1(\cdot,\cdot)$ is the generalized Marcum $Q$ function \cite{cantrell1987comparison}.
Furthermore, the probability of correctly identifying the AL SC-FDMA signal for certain channel coefficients can be expressed as

\vspace{-0.35cm}

\begin{equation}
\begin{array}{l}
\hspace{-0.2cm}P(\zeta = \textrm{AL} | \textrm{AL}, \boldsymbol{h})=1-[ (1-e^{-\gamma/2})^{N+\nu-6L_h}
 \prod_{\tau \in \Lambda} (1-Q_1 (\sqrt{\mathcal{P}_\tau}, \sqrt{\gamma})) ].
\end{array}\label{eq:P_AL_th_ch4}
\end{equation}

Then, the probability of correctly identifying the AL SC-FDMA signal can be calculated as $P(\zeta = \textrm{AL} | \textrm{AL})= \int_{\boldsymbol{h}} P(\zeta \hspace{-0.25cm}= \hspace{-0.25cm}\textrm{AL} | \textrm{AL}, \boldsymbol{h}) p(\boldsymbol{h}) d \boldsymbol{h}$, where $p(\boldsymbol{h})$ is the probability density function of $\boldsymbol{h}$. Finally,
 the average probability of correct identification is defined as
$P_c=0.5$ $\left[P(\zeta = \textrm{AL} | \textrm{AL})+P(\zeta = \textrm{SM} | \textrm{SM})\right]$ when AL and SM SC-FDMA are considered to be received with equal probability.

\vspace{-0.25cm}
\section{Simulation results \label{sec:Simulation_ch4}}

The evaluation of the proposed identification algorithm was done by considering the LTE SC-FDMA signal with a useful duration of 66.66 $\mu$sec and subcarrier spacing of 15 kHz. Unless otherwise mentioned, QPSK modulation, $N=512$ (5 MHz double sided bandwidth), $\nu=N/8$, $\lambda=2$, $N_B=1000$, and $P_{f}=10^{-3}$ were used. The received signal was affected by  frequency-selective Rayleigh fading channel with $L_h=3$ statistically independent taps\footnote{ Note that the identification performance depends on $L_h$ and usually degrades as $L_h$
increases, especially at low SNRs. In this case, using a larger number of received blocks can alleviate the degradation.} and an exponential power delay profile, $\sigma^{2}(l)=\exp(-l/5)$, where $l=0,...,L_h-1$. The out-of-band noise was removed at the receive-side with a Butterworth filter, and the signal-to-noise ratio (SNR) was considered at the output of this filter. The average probability of correct identification, $P_c$, was mainly showed as a measure of the identification performance. This was calculated based on 1000 trials for each signal.

Fig. \ref{fig:NB_ch4} (a) shows  the analytical and simulation results for the probability  of correct identification of AL and SM SC-FDMA signals achieved with the proposed algorithm  for various numbers of subcarriers, $N$.
In general, the theoretical findings are in good agreement with the simulation results. A
 slight difference  can be observed for $N=256$ when compared with larger values of $N$; this is because of the difference between the estimated variance in the denominator of (\ref{eq:Test_function_ch4}) and the actual value shown in (\ref{eq:var_th_ch4}). Indeed, this difference decreases as $N$ increases,
 as a better estimate of the variance, $\sigma^2_\epsilon$ is achieved in such cases. 
On the other hand, it can be seen that the identification performance significantly improves as  $N$ increases, as the peak values in $\mathcal{A}^{\textrm{AL}}_r(\tau) $ are proportional to $(N+\nu)$. 

\vspace{0.15cm}
The effect of the modulation format on  $P_c$ is illustrated in Fig. \ref{fig:NB_ch4} (b) with $N_B=2000$. As can be seen, the same performance is achieved for the PSK modulations, i.e., QPSK and 8-PSK,  while a performance degradation  occurs for 16-QAM and 64-QAM modulations. The dependence of the proposed algorithm on the modulation format can be explained, as the values of the discriminating peaks in 
 $\mathcal{A}_r^{\textrm{AL}} (\tau)$ are directly proportional to $\kappa_{d,4,2}$.  Theoretical values of $\kappa_{d,4,2}$ are given in \cite{dobre2007survey} for various unit variance constellations, and $\kappa_{d,4,2}$ equals $-1$ for QPSK and 8-PSK, and -0.68 and -0.619 for 16-QAM and 64-QAM, respectively \cite{dobre2007survey}. 

\vspace{0.15cm}
Fig. \ref{fig:NB_ch4} (c) presents  $P_c$  for various numbers of SC-FDMA blocks, $N_B$, with the interleaved ($\lambda=2$) and localized ($\lambda=512/312$) subcarrier mapping. One can notice that  the identification performance is not affected by the mapping, as the discriminating peaks still occur. 
Furthermore, the performance of the proposed algorithm enhances with $N_B$, as  more accurate estimates of the discriminating peaks in $\mathcal{A}^{\textrm{AL}}_r(\tau)$  are achieved.

\vspace{0.15cm} 
 Table \ref{tab:my-label}  presents the effect of the timing offset, $\mu$, frequency offset, $\Delta f$, and Doppler frequency, $|f_d|$, on the average probability of correct identification, $P_c$ for SNR=5, 10 dB. Note that the timing offset is normalized to the sampling interval \cite{choqueuse2008hierarchical}, whereas the frequency offset and Doppler frequency are normalized to the subcarrier spacing.   A slight degradation in the identification performance is observed at SNR=5 dB when $\mu$ increases. However, the performance is basically not affected by the timing offset at SNR= 10 dB. This is because the effect of timing offset can be modeled as an additional noise component that affects the discriminating peaks in $\mathcal{A}_r^{\textrm{AL}} (\tau)$; the effect of such a noise component vanishes at high SNRs. 
It should be noted that when the timing offset is a multiple integer of the sampling interval, the discriminating peaks in $\mathcal{A}_r^{\textrm{AL}}(\tau)$ are shifted by  that offset, which does not affect the performance (results are not included here due to the space consideration). On the other hand,  the performance drops for a normalized frequency offset
beyond $10^{-4}$. It should be mentioned that although the frequency offset does not affect
the identification of SM SC-FDMA, it affects the identification of the AL SC-FDMA
signal, as the discriminating peaks reduce when increasing the frequency offset. Hence,
the algorithm requires accurate estimation and compensation of the carrier frequency
recovery. A similar behavior is observed for  $|f_d|$; a good performance is achieved for  pedestrian mobility speeds.

\vspace{0.2cm}


\vspace{-0.55cm}
\section{Conclusion \label{sec:conclusion_ch4}}

The identification of SM SC-FDMA and AL SC-FDMA signals was addressed in this paper for the single receive antenna scenario. A fourth-order statistic of the received signal was employed as a discriminating feature. It was shown that the proposed  statistic exhibits statistically significant peaks for the AL SC-FDMA signal, while it does not for the SM SC-FDMA signal. Furthermore, a constant false alarm rate criterion was developed for decision-making. Analytical results for the identification performance of the proposed algorithm were derived. It was shown that simulation and theoretical findings match. The applicability of the proposed algorithm was proved through extensive simulations; this does not require estimation of noise power or channel coefficients, block timing synchronization,  and modulation identification. Furthermore, it is robust to timing offset, while requiring carrier frequency recovery and being suitable for pedestrian mobility~speeds. 
In future work, we plan to explore other test statistics, which are also applicable to identify STBCs with binary PSK, and are robust to the carrier frequency offset and mobility speeds.

\begin{table}[]
\vspace{0.55cm}
\centering
\caption{$P_c$ for different values of the normalized timing offset, $\mu$; normalized frequency offset, $\Delta f$; and normalized Doppler frequency, $|f_d|$.}
\label{tab:my-label}
\begin{tabular}{ll|c|c|c|c|c|c|}
\cline{3-8}
                                                &    & \multicolumn{2}{c|}{$\mu$} & \multicolumn{2}{c|}{$\Delta f$} & \multicolumn{2}{c|}{$|f_d|$} \\ \cline{3-8} 
                                                &    & $0.25$       & $0.5$       & $10^{-4}$      & $5 \times 10^{-4}$     & $10^{-4}$     & $10^{-3}$    \\ \hline
\multicolumn{1}{|c|}{\multirow{2}{*}{SNR (dB)}} & 10 & 1            & 0.98        & 0.98           & 0.73           & 0.99          & 0.97         \\ \cline{2-8} 
\multicolumn{1}{|c|}{}                          & 5  & 0.94         & 0.92        & 0.96           & 0.63           & 0.98          & 0.93         \\ \hline
\end{tabular}
\end{table}

\vspace{-0.75cm}
\bibliographystyle{IEEEtran}
\bibliography{IEEEabrv,My_Ref_bib}
\end{document}